\documentclass[12pt,fullpage]{article}
\usepackage[english]{babel}
\usepackage{epsfig,epsf}
\usepackage{graphicx}
\usepackage{hyphenat}
\author{V.A. Abramovsky, N.V. Prikhod'ko \\
Novgorod State University, B. S.-Peterburgskaya Street 41, \\
Novgorod the Great, Russia, 117259}
\title{Dipole amplitude correlation in saturation model beyond mean field approximation}

\begin{document}
\maketitle

\abstract{In this paper we calculate dipole amplitude and dipole amplitude correlations using first and
second equation from Balitsky-Kovchegov hierarchy.
Our analysis shows that even in presence of weak dipole correlation in initial condition mean field approximation
breaks down through evolution. This difference asymptotically grows not only by absolute value but also 
by relative difference of dipole correlation from mean field value.
This could affect physical values related to dipole correlation such as elliptic flow $v_2$ and 'back-forward' asymmetry in saturation model.
}

\section{Introduction}

It is well known that first equation in Balitsky-Kovchegov hierarchy \cite{Balitsky:1995ub,Kovchegov:1999yj} in mean field approximation have the following form
\begin{equation}
\partial_Y \langle T(r,Y)\rangle=\frac{\bar\alpha}{2\pi}\int d^2 z \frac{r^2}{z^2(r\!-\!z)^2}
(\langle T(z,Y)\rangle+\langle T(r\!-\!z,Y)\rangle-\langle T(r,Y)\rangle-\langle T(z,Y)\rangle \langle T(r\!-\!z,Y)\rangle)
\label{balitsky_eq1_mf}
\end{equation}

where $\langle T(r,Y)\rangle$ is dipole scattering amplitude.

As was proposed in \cite{Munier:2003vc} this equation can be gradually simplified with the following transformation
\begin{equation}
T(k,Y)=\int_0^{\infty}\frac{dr}{r}J_0(k r)T(r,Y)
\label{rk_trans}
\end{equation}

Which lead to simplified form 
\begin{equation}
\partial_{\bar\alpha Y} \langle T(L,Y)\rangle=\chi(-\partial_L) \langle T(L,Y)\rangle-
{\langle T(L,Y)\rangle}^2
\end{equation}

Where $L=ln(k^2)$ and integral operator $\chi$ is defined by following equation

\begin{equation}
\chi(-\partial_L) T(L)=\int_{-\infty}^{+\infty}dL^\prime
\left(\frac{T(L^\prime)-e^{L-L^\prime}T(L)}{\left|
1-e^{L-L^\prime}\right|}+
\frac{e^{L-L^\prime}T(L)}
{\sqrt{4+e^{2(L-L^\prime)}}}
\right)\
\label{BK_kernel_L}
\end{equation}

\begin{figure}
\resizebox{1\hsize}{!}{\epsffile{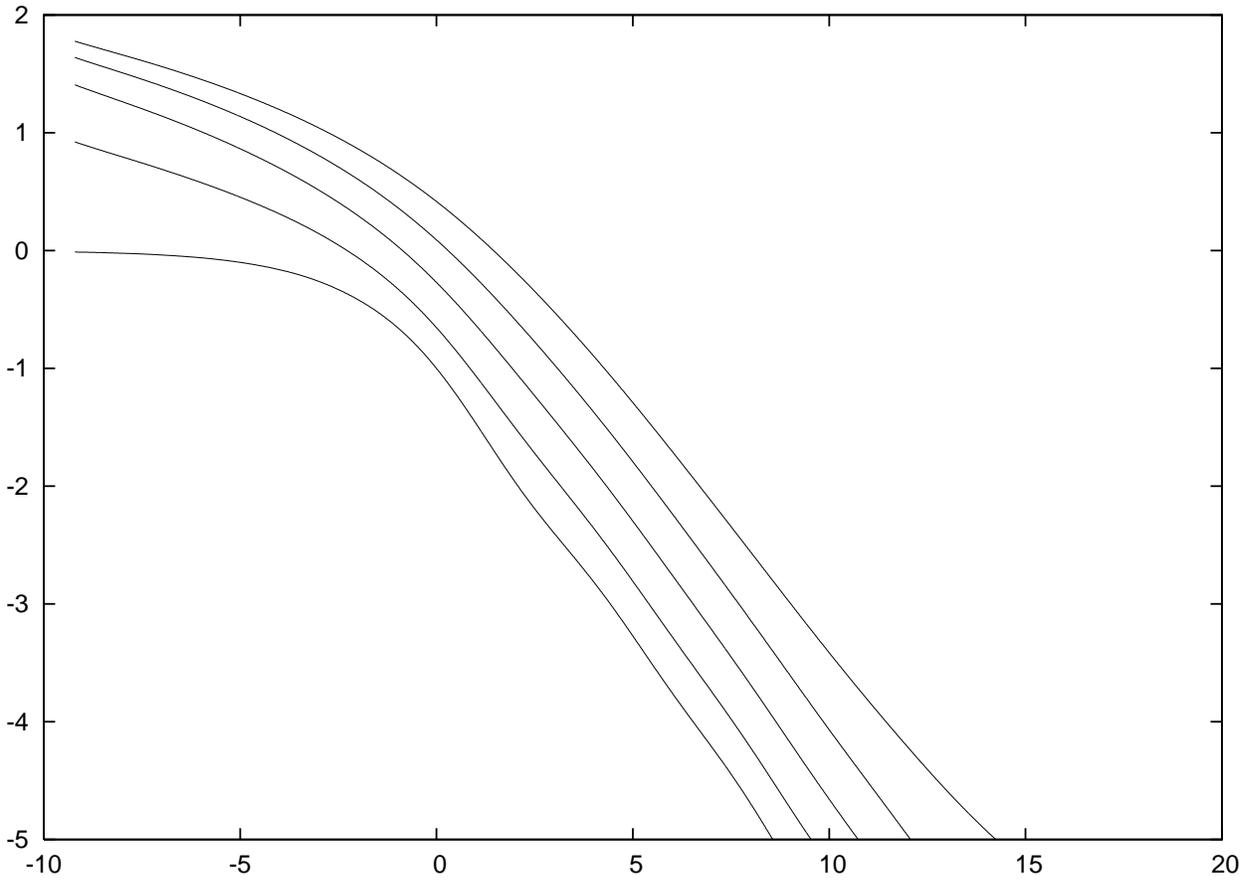}}
\caption{Evolution of dipole scattering amplitude defined by Balitsky-Kovchegov equation. $log(N(k,Y))$ from $log(k^2)$
dependence for $Y=0,1,2,3,4$ from left to right}
\label{fig_BK_1}
\end{figure}

Solution of equation (\ref{balitsky_eq1_mf}) was carefully examined by many authors using numerical simulation.
Moreover as was showed in \cite{Munier:2003vc} this equation belongs to Fisher-Kolmogorov-Petrovsky-Piscounov (FKPP) equation class \cite{Fisher}. Solution of this equation has quite remarkable properties. For physically acceptable initial condition
at asymptotically large $Y$, dipole scattering amplitude $\langle T(k,Y)\rangle$ can be represented as traveling wave in $k$ space
with $Y$ as time (fig. \ref{fig_BK_1}).
Moreover at asymptotically high $Y$ traveling wave form does not depend on initial condition.

However it is not clear if this holds true in case where mean field approximation of initial condition is not valid.
Since there is no method for analytical solution yet the only way to test it is through numerical simulation.
Let's consider first and second Balitsky-Kovchegov equations.

\begin{equation}
\partial_Y \langle T(r,Y)\rangle=\frac{\bar\alpha}{2\pi}\int d^2 z \frac{r^2}{z^2(r\!-\!z)^2}
(\langle T(z,Y)\rangle+\langle T(r\!-\!z,Y)\rangle-\langle T(r,Y)\rangle-\langle T(z,Y)T(r\!-\!z,Y)\rangle)
\label{balitsky_eq1}
\end{equation}

\begin{eqnarray}
\label{balitsky_eq2}
\partial_Y \langle T(r_1,Y)T(r_2,Y)\rangle
=\frac{\bar\alpha}{2\pi}\int d^2 z \frac{r_1^2}{z^2(r_1\!-\!z)^2}
(\langle T(z,Y)T(r_2,Y)\rangle+\langle T(r_1\!-\!z,Y)T(r_2,Y)\rangle-\\\nonumber
\langle T(r_1,Y)T(r_2,Y)\rangle-\langle T(z,Y)T(r_1\!-\!z,Y)T(r_2,Y) \rangle )\\\nonumber
+\frac{\bar\alpha}{2\pi}\int d^2 z \frac{r_2^2}{z^2(r_2\!-\!z)^2}
(\langle T(r_1,Y)T(z,Y)\rangle+\langle T(r_1,Y)T(r_2\!-\!z,Y)\rangle-\\\nonumber
\langle T(r_1,Y)T(r_2,Y)\rangle-\langle T(r_1,Y)T(z,Y)T(r_2\!-\!z,Y) \rangle )\ ,
\end{eqnarray}

Using (\ref{rk_trans}), simplified form can be obtained

\begin{equation}
\partial_{\bar\alpha Y} \langle T(L,Y)\rangle=\chi(-\partial_L) \langle T(L,Y)\rangle-
\langle T^2(L,Y)\rangle
\end{equation}
\begin{eqnarray}
\label{balitsky_eqk}
\partial_{\bar\alpha Y} \langle T(L_1,Y)T(L_2,Y)\rangle=(\chi(-\partial_{L_1})+\chi(-\partial_{L_2}))
\langle T(L_1,Y)T(L_2,Y)\rangle\\\nonumber
-\langle T^2(L_1,Y)T(L_2,Y)\rangle-\langle T(L_1,Y)T^2(L_2,Y)\rangle
\end{eqnarray}

This pair of equations allows mean field solution like:
\begin{equation}
\langle T(L_1,Y)\cdots T(L_n,Y)\rangle=
\lambda^{n-1}\langle T(L_1,Y)\rangle\cdots \langle T(L_n,Y)\rangle
\label{factorized_sol}
\end{equation}

Beyond this approximation this equation should be solved numerically.

\section{Numerical solution method}
Is is clearly seen that (\ref{balitsky_eq1},\ref{balitsky_eq2}) is incomplete since it contain term $\langle T(L_1,Y)T(L_2,Y)T(L_3,Y)\rangle$, to make this system complete we should make some approximation. Let's suppose that
\begin{eqnarray}
\label{mean_2}
\langle T(L_1,Y)T(L_2,Y)T(L_3,Y)\rangle=\langle T(L_1,Y)\rangle \langle T(L_2,Y)\rangle \langle T(L_3,Y)\rangle+\\\nonumber
(\langle T(L_1,Y)T(L_2,Y)-\langle T(L_1,Y)\rangle \langle T(L_2,Y)\rangle) \langle T(L_3,Y)\rangle\\\nonumber
(\langle T(L_2,Y)T(L_3,Y)-\langle T(L_2,Y)\rangle \langle T(L_3,Y)\rangle) \langle T(L_1,Y)\rangle\\\nonumber
(\langle T(L_3,Y)T(L_1,Y)-\langle T(L_3,Y)\rangle \langle T(L_1,Y)\rangle) \langle T(L_2,Y)\rangle
\end{eqnarray}
To obtain numerical solution for (\ref{balitsky_eq1},\ref{balitsky_eq2}) we should first apply some regularization
scheme to (\ref{BK_kernel_L}).
The most oblivious way for this is cut infinite interval from both sides, ultraviolet and infrared.
Therefore lets regularize (\ref{BK_kernel_L}) as:

\begin{equation}
\chi(-\partial_L) T(L)=\int_{L_{min}}^{L_{max}}dL^\prime
\left(\frac{T(L^\prime)-e^{L-L^\prime}T(L)}{\left|
1-e^{L-L^\prime}\right|}+
\frac{e^{L-L^\prime}T(L)}
{\sqrt{4+e^{2(L-L^\prime)}}}
\right)\
\label{BK_kernel_L_reg}
\end{equation}

where $L_{min}<<0<<L_{max}$.
For this regularization it can be shown that residue can be safely dropped from numerical computation.

Lets choose another variable $x=\frac{L-L_0}{\delta L}$, where
$L_0=\frac{L_{min}+L_{max}}{2}$ and $\delta L=\frac{L_{min}-L_{max}}{2}$

Therefore we have for kernel:
\begin{equation}
\chi(x) T(x)=\int_{-1}^{1}\delta L dx^\prime
\left(\frac{T(x^\prime)-e^{(x-x^\prime)\delta L}T(x)}{\left|
1-e^{(x-x^\prime)\delta L}\right|}+
\frac{e^{(x-x^\prime)\delta L}T(x)}
{\sqrt{4+e^{2(x-x^\prime)\delta L}}}
\right)\
\label{BK_kernel_x}
\end{equation}

Defining $N_1(x,Y)=\langle T(x,Y) \rangle$ and $N_2(x_1,x_2,Y)=\langle T(x_1,Y)T(x_2,Y)\rangle-N_1(x_1,Y)N_2(x_2,Y)$
and using (\ref{mean_2}) equation (\ref{balitsky_eqk}) can be rewritten as

\begin{equation}
\partial_{\bar\alpha Y} N_1(x,Y)=\chi(x) N_1(x,Y)- N_1^2(x,Y)-N_2(x,x,Y)
\label{balitsky_eqx_1}
\end{equation}
\begin{eqnarray}
\label{balitsky_eqx_2}
\partial_{\bar\alpha Y} N_2(x_1,x_2,Y)\rangle=(\chi(x_1)+\chi(x_2))
N_2(x_1,x_2,Y)\\\nonumber
-2N_2(x_1,x_2,Y)N_1(x_1,Y)-2N_2(x_1,x_2,Y)N_1(x_2,Y)
\end{eqnarray}

Now there is oblivious way to solve this equation. Lets approximate $N_1(x,Y)$ and $N_2(x_1,x_2,Y)$ with

\begin{equation}
N_1(x,Y)=\sum_{0}^{N_{max}}{C^i_1(Y)U_i(x)}
\end{equation}

\begin{equation}
N_2(x_1,x_2,Y)=\sum_{i=0}^{N_{max}}\sum_{j=0}^{N_{max}}{C^{ij}_2(Y)U_i(x_1)U_j(x_2)}
\end{equation}

where $U_i(x)$ is some set of orthogonal functions with
\begin{equation}
\int_{-1}^{1}{U_i(x)U_j(x)\mu(x)dx=\delta_{ij}\mu_{i}}
\end{equation}

With this equation (\ref{balitsky_eqx_1},\ref{balitsky_eqx_2}) can be rewritten as

\begin{equation}
\partial_{\bar\alpha Y} C^i_1(Y)=\frac{1}{\mu_i}\chi_{ij}C_1^j(Y)- \frac{1}{\mu_i}F_{ijk}C_2^{jk}(Y))
\label{balitsky_eq_C1}
\end{equation}

\begin{eqnarray}
\label{balitsky_eq_C2}
\partial_{\bar\alpha Y} C_2^{ij}(Y)=\frac{1}{\mu_i}\chi_{ik}C_2^{kj}(Y)+\frac{1}{\mu_j}\chi_{jk}C_2^{ik}(Y)-\\\nonumber
-\frac{2}{\mu_i} C^i_1 C^{kl} F_{klj}-\frac{2}{\mu_j} C^j_1 C^{kl} F_{kli}
\end{eqnarray}

where
\begin{equation}
\chi_{ij}=\int_{-1}^{1}{U_i(x)\chi(x)U_j(x)\mu(x)dx}
\end{equation}

\begin{equation}
F_{ijk}=\int_{-1}^{1}{U_i(x)U_j(x)U_k(x)\mu(x)dx}
\label{F0}
\end{equation}

System of ordinary differential equation (\ref{balitsky_eq_C1},\ref{balitsky_eq_C2}) can easily be solved by Runge-Kutta-Fehlberg method.

The most oblivious choose for orthogonal basis $U_i(x)$ is the first kind Chebyshev polynomials. However in this case solution is unstable.
The instability comes from points $|x|=1$. Same applies for second kind Chebyshev polynomials.
Therefore we choose Legendre polynomials as orthogonal basis for our purposes.

\section{Results}
For numerical computation we set $N_{max}=64$, $L_{min}=-10$, $L_{max}=50$, $\alpha_s=0.2$.
\begin{eqnarray}
N_1(k,0)=\frac{e^{-k/Q_s-(k/Q_s)^2}+1}{1+k}\\\nonumber
N_2(k_1,k_2,0)=\epsilon N_1(k_1,Y)N_1(k_2,Y)exp(-log^2(k_1/k_2))
\end{eqnarray}
with $\epsilon=10^{-3}$.
Numerical solution for absolute and relative correlation $N_2(k_1,k_2,Y)$ is shown on fig. 2 and fig. 3 respectively. 

\begin{figure}
\resizebox{1\hsize}{!}{\epsffile{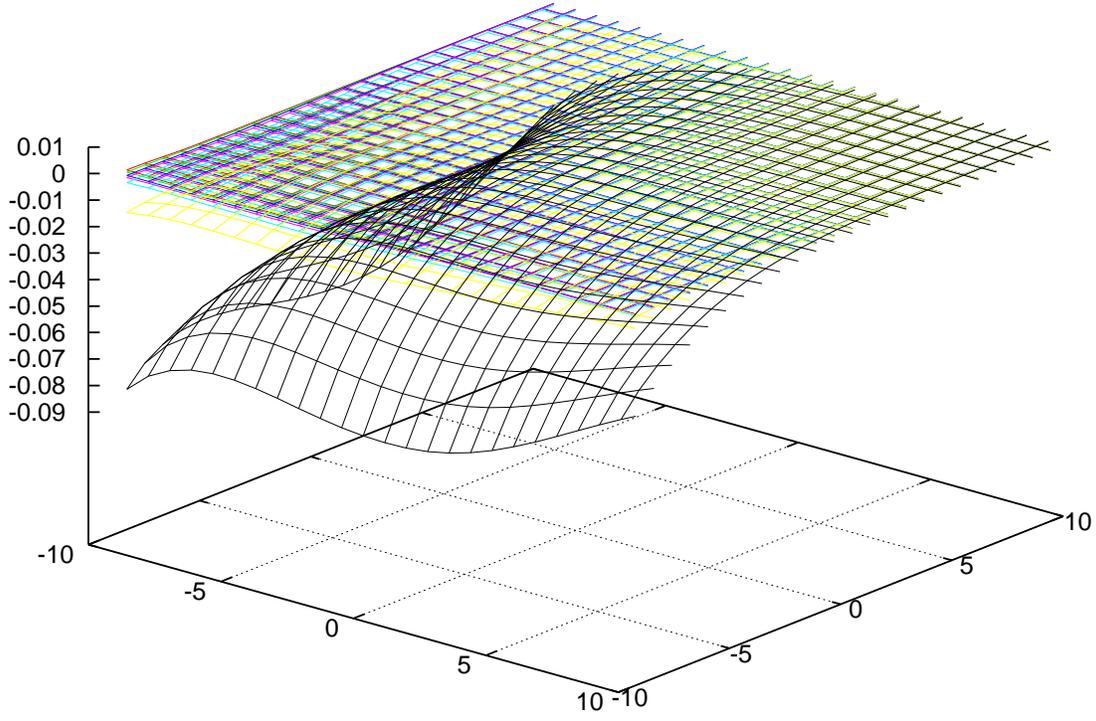}}
\caption{Evolution of dipole correlation defined by Balitsky-Kovchegov equation.
$N_2(k_1,k_2,Y)$ from $log(k_1^2), log(k_2^2)$ dependence for $Y=0,1,2,3,4,5$ from up to down}
\label{fig_BK_2}
\end{figure}
\begin{figure}
\resizebox{1\hsize}{!}{\epsffile{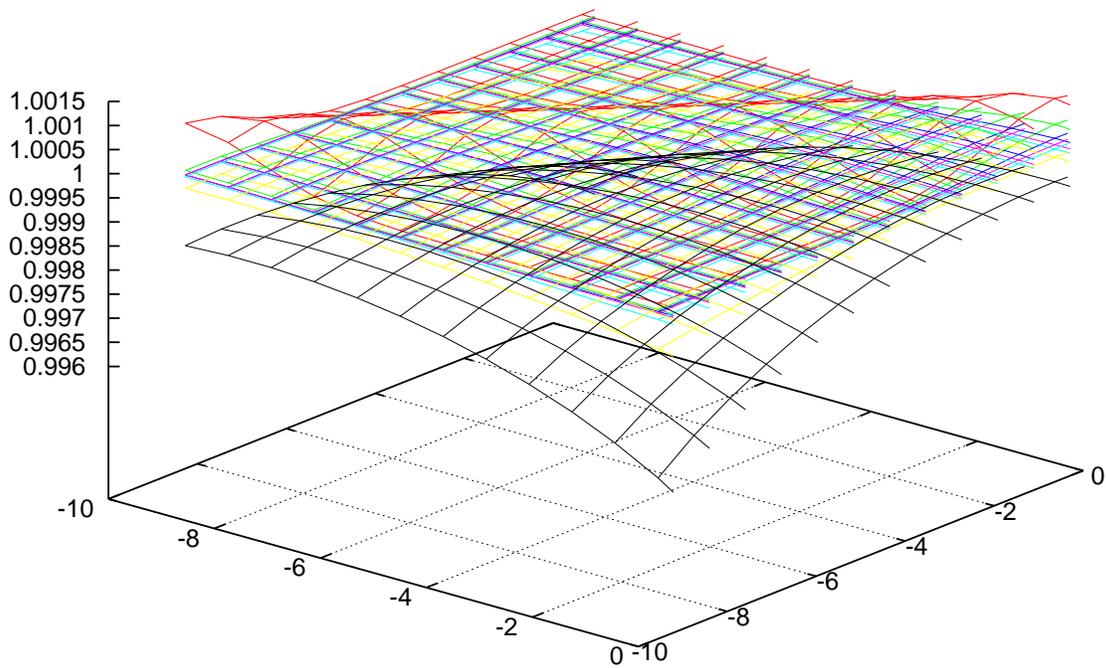}}
\caption{Evolution of dipole correlation defined by Balitsky-Kovchegov equation.
$N_2(k_1,k_2,Y)/N_1(k_1,Y)N_1(k_2,Y)+1$ from $log(k_1^2), log(k_2^2)$ dependence for $Y=0,1,2,3,4,5$ from up to down}
\label{fig_BK_2}
\end{figure}

It is clearly seen that $\langle T(k_1,Y)T(k_2,Y) \rangle$ is smaller what value $\langle T(k_1,Y) \rangle \langle T(k_2,Y) \rangle$ supposed by mean field approximation. Moreover not only absolute difference between $\langle T(k_1,Y)T(k_2,Y) \rangle$ and $\langle T(k_1,Y) \rangle \langle T(k_2,Y) \rangle$ does not vanish with $Y$ but so does relative difference.
This difference affect physical values related to dipole correlation such as elliptic flow $v_2$ and 'back-forward' asymmetry in saturation model. Recently Balitsky-Kovchegov hierarchy was reformulated \cite{Iancu:2004iy} using stochastic field theory. It was showed what (\ref{balitsky_eq2}) should contain additional 'border' term which related to fluctuation. It therefore possible that $N_2(k_1,k_2,Y)$ behaviour will be changed drastically. This required further investigation.

\section*{Acknowledgments}
This work was supported by grants RFBR 03-02-16157-a, RFBR 05-02-08266-ofi\_a.

\end{document}